\documentclass[11pt]{article}
\usepackage{blois,epsfig}

\def\be{\begin{equation}}
\def\ee{\end{equation}}
\def\bea{\begin{eqnarray}}
\def\eea{\end{eqnarray}}

\newcommand{\simlt}
{\mbox{\raisebox{-0.5ex}{$\textstyle \; \sim$}
\raisebox{ 0.8ex}{$\textstyle  \!\!\!\!\!\!\! <$  }}}
\begin{document}
\vspace*{4cm}
\title{SUPERMASSIVE FERMION BALLS AND CONSTRAINTS
FROM STELLAR DYNAMICS NEAR Sgr A$^{*}$}
\author{NEVEN BILI\'{C}$^{a,b}$,
GARY B.\ TUPPER$^a$, and RAOUL D.\ VIOLLIER$^a$
}
\address{$^a$Institute of Theoretical Physics and Astrophysics,
 Department of Physics, University of Cape Town,
 Private Bag, Rondebosch 7701, South Africa \\
Email: viollier@physci.uct.ac.za \\
$^b$Rudjer Bo\v skovi\'c Institute,
P.O. Box 180, 10002 Zagreb, Croatia \\
Email: bilic@thphys.irb.hr}

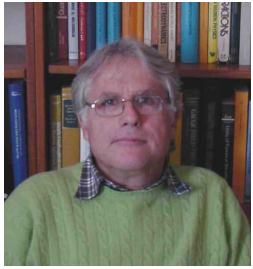
\maketitle

\abstracts{
All presently known stellar-dynamical constraints
on the
size and mass of the supermassive compact dark object
at the Galactic center are consistent with a ball of
self-gravitating, nearly non-interacting, degenerate
fermions with mass between 76 and 491 keV/$c^{2}$,
for a degeneracy factor $g$ = 2.
Sterile neutrinos of 76 keV/$c^{2}$ mass, which are
mixed with at least one of the active neutrinos with
a mixing angle $\theta \sim 10^{-7}$, are produced in
about the right amount in the early Universe and
may be responsible for the formation of the
supermassive degenerate fermion balls and black
holes at the galactic centers via gravitational
cooling.}
\section{Introduction}
In a recent paper Sch\"{o}del {\it et al} reported
a new set of
data \cite{schod1}
including the corrected old
measurements \cite{eck2}
on the projected positions of the star S2(S0-2)
that was observed during the last decade with the
ESO telescopes in La Silla (Chile). The combined
data suggest that S2 (S0-2)is moving on a
Keplerian orbit with a period of 15.2 yr around
the enigmatic strong radiosource Sgr A$^{*}$ that
is widely believed to be a black hole with a mass
of about 2.6 $\times$ 10$^{6} M_{\odot}$
\cite{eck2,ghez3}.
The salient feature of the new adaptive optics
data is that, between April and May 2002,
S2(S0-2) apparently sped past the point of
closest approach with a velocity $v$ $\sim$
6000 km/s at a distance of about 17 light-hours
\cite{schod1} or 123 AU from Sgr A$^{*}$.
Another star, S0-16 (S14), which was observed
during the last few years by Ghez {\it et al}
\cite{ghez4}
with the Keck telescope in Hawaii, made
recently a spectacular U-turn, crossing the
point of closest approach at an even smaller
distance of 8.32 light-hours or 60 AU from
Sgr A$^{*}$ with a velocity $v$ $\sim$
9000 km/s. Ghez {\it et al} \cite{ghez4} thus conclude
that the gravitational potential around
Sgr A$^{*}$ has approximately $r^{-1}$ form,
for radii larger than 60 AU, corresponding to
1169 Schwarzschild radii of 26 light-seconds
or 0.051 AU for a 2.6
$\times$ 10$^{6} M_{\odot}$ black hole.
Although the baryonic alternatives are
presumably ruled out, this still leaves
some room for the interpretation of the
supermassive compact dark object at the
Galactic center in terms of a finite-size
non-baryonic dark matter object rather
than a black hole. In fact, the supermassive
black hole
paradigm may eventually only be proven
or ruled out by comparing it with
credible alternatives in terms of
finite-size non-baryonic objects
\cite{mun9}.

The purpose of this paper is to explore,
using the example of a sterile neutrino
as the dark matter particle candidate,
the implications of the recent observations
for the degenerate fermion ball scenario of
the supermassive compact dark objects which
was developed during the last decade
\cite{mun9,viol5,viol6,bil7,bil8,mun10,bil11}.

\section{Stellar-dynamical constraints for fermion balls}
In a self-gravitating ball of degenerate
fermionic matter, the gravitational
pressure is balanced by the degeneracy
pressure of the fermions due to the
Pauli exclusion principle.
Nonrelativistically, this scenario is
described by the Lane-Emden equation
with polytropic index $p = 3/2$. Thus
the radius $R$ and mass $M$ of a ball
of self-gravitating, nearly non-interacting
degenerate fermions of mass $m$ and
degeneracy $g$ scale as \cite{viol6}
\begin{eqnarray}
R=\left[\frac{91.869
\hbar^{6}}{m^{8} G^{3}}\left( \frac{2}{g}
\right)^{2} \frac{1}{M} \right]^{1/3}
\hspace*{- 0.02cm}=3610.66\;\mbox{ld}
\left(\frac{15\;\mbox{keV}}{mc^{2}}
\right)^{8/3} \left(\frac{2}{g} \right)^{2/3}
\left( \frac{M_{\odot}}{M} \right)^{1/3}.
\end{eqnarray}
Here 1.19129 ld = 1 mpc = 206.265 AU.
Using the canonical value
$M = 2.6 \times 10^{6} M_{\odot}$
and $R$ $\leq$ 60 AU for the supermassive
compact dark object at the Galactic center,
we obtain a minimal fermion mass of
$m_{\rm min}$ = 76.0 keV/$c^{2}$ for
$g$ = 2, or $m_{\rm min}$ = 63.9 keV/$c^{2}$
for $g$ = 4.

The maximal mass for a degenerate
fermion ball, calculated in a general
relativistic framework based on the
Tolman-Oppenheimer-Volkoff equations,
is the Oppenheimer-Volkoff (OV) limit
\cite{bil7}
\begin{equation}
M_{\rm OV} = 0.38322 \;
\frac{M_{\rm Pl}^{3}}{m^{2}} \,
\left( \frac{2}{g} \right)^{1/2}
=  2.7821 \times 10^{9} M_{\odot} \,
\left( \frac{15 \; \mbox{keV}}{mc^{2}}
\right)^{2} \, \left( \frac{2}{g}
\right)^{1/2}  ,
\end{equation}
where
$M_{\rm Pl} = (\hbar c/G)^{1/2} = 1.2210
\times 10^{19}$ GeV is the Planck mass.
Thus, for $m_{\rm min}$ = 76.0 keV/$c^{2}$
and $g$ = 2, or $m_{\rm min}$ = 63.9 keV/$c^{2}$
and $g$ = 4, we obtain
$M_{\rm OV}^{\rm max} = 1.083 \times 10^{8}
M_{\odot}$.
In this scenario all supermassive compact
dark objects with mass $M > M_{\rm OV}^{\rm max}$
must be black holes, while those with
$M \leq M_{\rm OV}^{\rm max}$ are fermion balls.
Choosing as the OV limit the canonical mass
of the compact dark object at the center of
the Galaxy,
$M_{\rm OV}^{\rm min}$ = 2.6 $\times$ 10$^{6}$
$M_{\odot}$,
yields a maximal fermion mass of
$m_{\rm max}$ = 491 keV/$c^{2}$
for $g$ = 2, or $m_{\rm max}$ =
413 keV/$c^{2}$ for $g$ = 4.

The masses of the supermassive compact
dark objects
discovered so far at the centers of both
active and
inactive galaxies are all in the range
\cite{korm12}
$10^{6} M_{\odot} \; \simlt \; M \; \simlt \;
3 \times 10^{9} M_{\odot}$.
Thus, as $M_{\rm OV}^{\rm max}$ falls into
this range as well, we need both supermassive
fermion balls
($M \leq M_{\rm OV}^{\rm max}$)
and black holes ($M > M_{\rm OV}^{\rm max}$)
to describe the observed phenomenology.
At first sight, such a hybrid scenario
does not
seem to be particularly attractive.
However, it
is important to note that a similar
scenario is
actually realized in Nature, with
the co-existence
of neutron stars
and stellar-mass black
holes
as observed
in stellar binary systems in the Galaxy
\cite{bland13}.
It is thus conceivable that Nature allows for the
co-existence of supermassive fermion balls and
black holes as well.
\section{Cosmological constraints for
sterile neutrino dark matter}
If the supermassive compact dark object at the
Galactic center is indeed a degenerate fermion
ball of mass
$M$ = 2.6 $\times$ 10$^{6} M_{\odot}$ and radius
$R$ $\leq$ 60 AU, the fermion mass must be in the
range
$76.0 \; \mbox{keV}/c^{2} \; \leq \; m \; \leq \;
491 \; \mbox{keV}/c^{2} \; \; \mbox{for}
\; \; g = 2$, or
$63.9 \; \mbox{keV}/c^{2} \; \leq \; m \;
\leq \; 413 \; \mbox{keV}/c^{2} \; \;
\mbox{for} \; \; g = 4$.
It would be most economical if this particle
could represent the dark matter particle of
the Universe, as well.
The conjectured fermion could be a sterile
neutrino $\nu_{s}$ which does not participate
in the weak interactions.
We will now assume that its mass and degeneracy
factor is
$m_{s}$ = 76.0 keV/$c^{2}$ and $g_{s}$ = 2,
corresponding
to the largest fermion ball that is consistent
with the stellar-dynamical constraints. In order
to make sure that this fermion is actually
produced in the early Universe it must be mixed
with at least one active neutrino, e.g., the
$\nu_{e}$. Indeed, for an electron neutrino
asymmetry
$L_{\nu_{e}} =
\frac{n_{\nu_{e}} - n_{\overline{\nu}_{e}}}
{n_{\gamma}} \sim 10^{-2}$
and a mixing angle $\theta_{es} \sim 10^{-7}$
\cite{abza14},
incoherent resonant and non-resonant active
neutrino scattering in the early Universe
produces sterile neutrino matter amounting
to the required fraction
$\Omega_{m} h^{2} = \left( 0.135_{- 0.009}^{+0.008}
\right)$ \cite{wmap15} of the critical density
of the Universe today.
An electron
neutrino asymmetry of $L_{\nu_{e}} \sim 10^{-2}$
is compatible with the observational limits
\cite{kang16,espo17}
which is constrained by
$- 4.1 \times 10^{-2} \; \leq \; L_{\nu_{e}}
\; \leq 0.79$.

It is interesting to note that
incoherent resonant scattering of active
neutrinos produces
quasi-degenerate sterile neutrino matter,
while incoherent non-resonant active neutrino
scattering yields sterile neutrino matter that
has approximately a thermal spectrum
\cite{abza14}. Quasi-degenerate sterile neutrino
matter may contribute towards the formation of
the supermassive compact dark objects at the
galactic centers, while thermal sterile
neutrino matter is mainly contributing to the
dark matter of the galactic halos.
In fact, it has been recently shown \cite{bil8}
that an extended cloud of degenerate fermionic
matter will eventually undergo gravitational
collapse and form a degenerate supermassive
fermion ball in a few free-fall times, if the
collapsed mass is below the OV limit. During
the formation, the binding energy of the nascent
fermion ball is released in the form of high-energy
ejecta at every bounce of the degenerate fermionic
matter through a mechanism similar to gravitational
cooling that is taking place in the formation of
degenerate boson stars  \cite{bil8}.
If the mass of the collapsed object is above
the OV limit, the collapse inevitably results
in a supermassive black hole.
\section{Observability of
degenerate sterile neutrino balls}
The mixing of the sterile neutrino with at
least one of the active neutrinos necessarily
causes
radiative decay of the sterile into an active
neutrino and a photon. The lifetime
is \cite{boeh18}
\begin{equation}
\tau \left( \nu_{s} \rightarrow \nu \gamma \right)
= \frac{8 \pi}{27 \; \alpha} \;
\frac{1}{\sin^{2} \theta_{es}} \;
\left( \frac{m_{\mu}}{m_{s}} \right)^{5} \;
\tau \; (\mu^{-} \rightarrow e^{-} +
\bar{\nu}_{e} + \nu_{\mu})
\end{equation}
yielding, for $\theta_{es} = 10^{-7}$ and
$m_{s}$ = 76.0 keV/$c^{2}$, a lifetime of
$\tau (\nu_{s} \rightarrow \nu \gamma) =
0.46 \times 10^{19}$ yr.
Although the X-ray luminosity due to the
radiative decay of diffuse sterile neutrino dark
matter in the Universe is presumably not observable,
because it is well below the X-ray background
radiation at this energy \cite{abza14}, it is
perhaps possible to detect such hard X-rays in
the case of sufficiently concentrated dark matter
objects.
In fact, this could be the smoking gun for both
the existence of the sterile neutrino and the
fermion balls. For instance, a ball of $M = 2.6
\times 10^{6} M_{\odot}$ consisting of degenerate
sterile neutrinos of mass
$m_{s}$ = 76.0 keV/$c^{2}$ \cite{mun10},
degeneracy factor $g_{s}$ = 2, and mixing angle
$\theta_{es}$ = 10$^{-7}$
would emit 38 keV photons with a luminosity
$L_{X} = 1.6 \times 10^{34} \mbox{erg/s}$
within a radius of 60 AU, 8.32 light hours or
7.6 $\times$ 10$^{-3}$ arcsec of Sgr A$^{*}$,
assumed to be at a distance of 8 kpc. The
current upper limit for X-ray emission from the
vicinity of Sgr A$^{*}$ is
$\nu L_{\nu} \sim 3 \times 10^{35}$ erg/s,
for an X-ray energy of $E_{X} \sim 60$ keV
\cite{maha19},
where $L_{\nu} = dL/d \nu$ is
the spectral luminosity. Thus the X-ray line at
38 keV could presumably only be detected if
either the angular or the energy resolution
or both, of the present X-ray detectors are
increased.

\end{document}